\documentclass[11pt,a4paper]{article}

\usepackage[T1]{fontenc}
\usepackage[utf8]{inputenc}
\usepackage[margin=2.2cm]{geometry}
\usepackage{amsmath,amssymb,amsthm,mathtools,bm}
\usepackage{graphicx}
\usepackage{booktabs}
\usepackage{microtype}
\usepackage{xcolor}
\usepackage[super,comma,sort&compress]{natbib}
\usepackage[colorlinks=true,linkcolor=blue!60!black,citecolor=blue!60!black,urlcolor=blue!60!black]{hyperref}
\usepackage[font=small,labelfont=bf]{caption}
\usepackage{authblk}

\renewcommand{\figurename}{Fig.}
\DeclareMathOperator{\sgn}{sgn}

\DeclareMathOperator{\Cov}{Cov}
\DeclareMathOperator{\Var}{Var}
\DeclareMathOperator{\diag}{diag}

\title{\textbf{Gauge-equivariant real-space renormalization of frustrated magnets and topological spin textures}}

\author[1]{Musen Li\thanks{These authors contributed equally to this work.}\textsuperscript{,}\thanks{Corresponding author: \texttt{musenli@shu.edu.cn}.}}
\author[1]{Tianhao Su\protect\footnotemark[1]}
\author[1]{Shunbo Hu}
\author[1]{Tong-Yi Zhang}

\affil[1]{Materials Genome Institute (MGI), Shanghai University, Shanghai, China}

\date{\today}

\begin{document}
\maketitle

\noindent\textbf{Real-space renormalization group transformations and equilibrium Monte Carlo sampling are foundational tools of statistical physics, yet both break down in frustrated and chiral magnets. Fixed geometric coarse-graining blocks average anti-aligned sublattices and erase emergent gauge fluxes on triangular, ruby, spin-glass and multi-shell lattices, whereas local Monte Carlo dynamics suffer catastrophic topological freezing between distinct winding-number sectors below the skyrmion-ordering transition. Here we show that decimating non-adjacent spins while preserving local bond-rotation rules defines an exact-likelihood, gauge-equivariant real-space renormalization flow that overcomes both barriers simultaneously. Through exact matrix coarsening of the coupled spin network, the flow extracts a gauge-invariant frustration deficit measuring destructive loop interference, separating unfrustrated order, spin liquids and spin glasses while recovering exact Wannier residual entropy. Paired with an annealed sequence of parallel heat-bath and energy-conserving over-relaxation sweeps, the method matches thermodynamic integration to $0.01\%$ on density-functional-calibrated $\mathrm{fcc\text{-}Pd/Fe/Ir(111)}$, maps its finite-temperature skyrmion melting boundary, images a $30\%$ softening of domain-wall magnon stiffness in centrosymmetric $\mathrm{Gd_2PdSi_3}$, and models experimental 219-atom ruby-lattice Rydberg spin-liquid snapshots, linking frustrated statistical mechanics to live spin-polarized quantum readout.}

\vspace{0.6em}

Real-space renormalization group (RG) transformations provide a direct microscope into collective behaviour across length scales, from Kadanoff block-spin decimation and Wilsonian scaling\cite{kadanoff1966scaling,wilson1975renormalization} through tensor renormalization group methods\cite{levin2007tensor,evenbly2015tensor} to modern machine-learning phase discovery and information-theoretic neural RG schemes\cite{carrasquilla2017machine,koch2018mutual,li2018neural,lenggenhager2020optimal}. Yet virtually all generative real-space RG constructions rely on fixed geometric $2\times 2$ or checkerboard decimations. On geometrically frustrated lattices hosting classical or quantum spin liquids (triangular, kagome and ruby)\cite{balents2010spin,savary2017quantum}, disordered $\pm J$ Edwards--Anderson spin glasses\cite{edwards1975theory,toulouse1977theory} and multi-shell chiral magnets, fixed geometric blocks average anti-aligned spins and erase the emergent gauge fluxes\cite{toulouse1977theory,kogut1979introduction} that govern the low-energy manifold. Autoregressive neural samplers\cite{wu2019solving,nicoli2020asymptotically} and neural-network quantum state reconstructions\cite{torlai2018neural} avoid geometric blocks on discrete graphs, but autoregressive models incur $O(N)$ sequential sampling latency and $O(N^2)$ parameter growth, and do not extend naturally to continuous $S^2$ spin textures. Conversely, continuous normalizing flows and Boltzmann generators\cite{noe2019boltzmann,rezende2020normalizing}, including gauge-equivariant flows for lattice field theories\cite{albergo2019flow,kanwar2020equivariant,boyda2021sampling}, typically operate at a single scale without an explicit frustration-resolving coarse-graining hierarchy.

A parallel barrier confronts equilibrium simulations of topological magnetism\cite{muhlbauer2009skyrmion,yu2010real,nagaosa2013topological,fert2017magnetic}. In real interfacial chiral magnets such as $\mathrm{fcc\text{-}Pd/Fe/Ir(111)}$\cite{vonmalottki2017enhanced,dupe2014tailoring,romming2013writing,romming2015field} and centrosymmetric frustrated skyrmion hosts such as $\mathrm{Gd_2PdSi_3}$\cite{kurumaji2019skyrmion,leonov2015multiply,okubo2012multiple}, competing exchange and anisotropic interactions stabilize quantized skyrmions ($Q \in \mathbb{Z}$) with internal helicity ($\varphi_0$) degrees of freedom, which have recently been proposed as macroscopic quantum logic elements\cite{psaroudaki2021skyrmion}. Below the skyrmion-ordering transition, local Markov chain Monte Carlo (MCMC), cluster updates\cite{swendsen1987nonuniversal,wolff1989collective}, parallel tempering (PT)\cite{hukushima1996exchange} and single-scale normalizing flows suffer severe topological freezing\cite{schaefer2011critical} across homotopy sectors $Q \to Q \pm 1$. Consequently, the exact finite-temperature free-energy landscape $\Delta F(Q, \varphi_0; T, B) = \Delta E - T\Delta S$ and the microscopic role of soft domain-wall magnons and entropic narrowing in skyrmion stability\cite{wild2017entropy,desplat2018thermal,bessarab2018lifetime} have remained difficult to quantify directly, while experimental spin-polarized scanning tunnelling microscopy (SP-STM)\cite{wiesendanger2009spin,wiesendanger2016nanoscale} and non-collinear magnetoresistance (NCMR)\cite{hanneken2015electrical} readout remain bottlenecked by sparse, single-axis projections and finite tip resolution.

Here we resolve both challenges through an exact-likelihood, gauge-equivariant real-space RG flow (GaugeRGFlow), summarized schematically in the conceptual overview below. Starting from a microscopic frustrated or chiral spin lattice $G^{(0)}$ (left), we encode each bond as a local spin-rotation rule ($O_{uv} \in \{\mathbb{Z}_2, SO(3)\}$) capturing antiferromagnetic sign flips or chiral canting; formally, these bond rules endow the lattice with a discrete principal $G$-bundle and a connection (sheaf) Laplacian\cite{singer2012vector,hansen2019toward} that interfaces directly with differentiable graph partitioning\cite{schuetz2022combinatorial,abate2025maxcutpool}. Across successive coarse-graining scales $G^{(0)} \to G^{(1)} \to \dots \to G^{(L)}$ (centre), we decimate a set of mutually non-adjacent spins $B_\ell$ (an independent set, $E_\ell(B_\ell) = \emptyset$) so that all spins in $B_\ell$ are integrated out simultaneously and exactly without destroying frustration. This strictly diagonalizes the eliminated block $\mathcal{L}_{B_\ell B_\ell}^{(\ell)}$, yielding exact $O(|B_\ell|)$ polar-projected Schur coarsening and a gauge-invariant frustration deficit $\Delta^{(\ell)} \ge 0$ measuring destructive loop interference. Reversing the hierarchy from infrared to ultraviolet (right) with $O(1)$ weight-shared parameters and $O(\log N)$ parallel depth lifts coarse topological sectors back to full microscopic spin configurations with exact analytical probabilities (zero stochastic divergence noise), enabling both unbiased free-energy sampling of chiral skyrmion textures and live active SP-STM readout. We validate the framework across frustrated spin glasses, density-functional-theory (DFT)-calibrated $\mathrm{fcc\text{-}Pd/Fe/Ir(111)}$ and $\mathrm{Gd_2PdSi_3}$, live millisecond SP-STM diagnostics and experimental 219-atom ruby-lattice Rydberg quantum-simulator data probing $\mathbb{Z}_2$ topological order\cite{semeghini2021probing,verresen2021prediction}. Companion applications to $(2+1)$D surface-code entropic maximum-likelihood decoding\cite{kitaev2003fault,dennis2002topological,bravyi2014efficient} and 1,024-qubit quantum approximate optimization readout mitigation are summarized in Extended Data Figs.~5--7 and reported separately.

\begin{figure}[!htbp]
  \centering
  \includegraphics[width=0.98\linewidth]{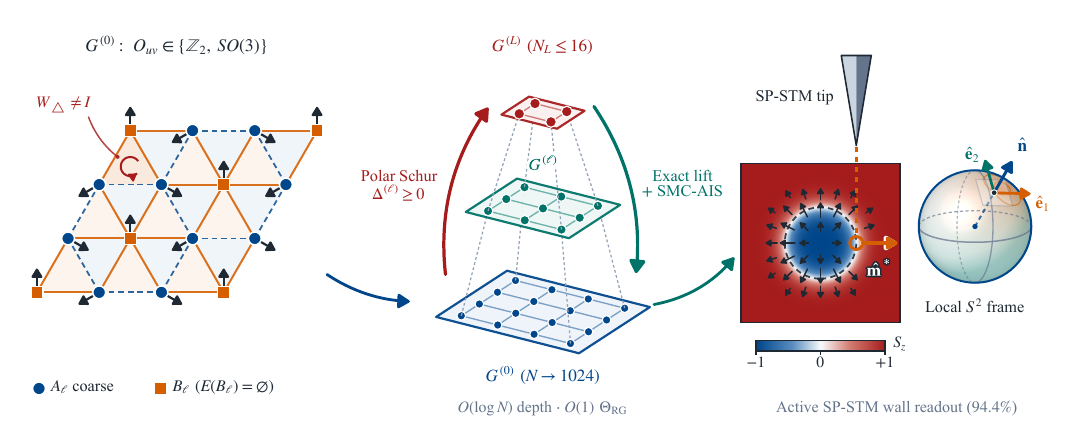}
  \caption*{\textbf{Conceptual overview (TOC).} Three-stage GaugeRGFlow pipeline described in the Introduction: microscopic frustrated spin bundle $G^{(0)}$ with bond-rotation rules $O_{uv}$ (left), multi-scale independent-set gauge decimation and polar-projected Schur coarsening $G^{(0)} \to G^{(1)} \to G^{(L)}$ extracting the gauge-invariant frustration deficit $\Delta^{(\ell)}$ (centre), and exact-likelihood IR-to-UV lifting to chiral skyrmion textures and active spin-polarized tunnelling readout (right).}
\end{figure}

\section*{Frustration as discrete curvature under coarse-graining}

Let $G^{(0)} = (V_0, E_0)$ be an arbitrary interaction graph with spins $\bm{S}_u \in \{-1, +1\}$ or $S^2 \subset \mathbb{R}^3$ governed by the multi-shell Hamiltonian
\begin{equation}
  H(\bm{S}) = \sum_{\{u,v\} \in E_0} \Big[ J_{uv}\, \bm{S}_u \cdot \bm{S}_v + \bm{D}_{uv} \cdot (\bm{S}_u \times \bm{S}_v) \Big] - K_z \sum_{u \in V_0} (S_u^z)^2 - \sum_{u \in V_0} \bm{h}_u \cdot \bm{S}_u,
  \label{eq:hamiltonian}
\end{equation}
where $J_{uv} < 0$ denotes ferromagnetic exchange, $J_{uv} > 0$ antiferromagnetic exchange, $\bm{D}_{vu} = -\bm{D}_{uv}$ is the antisymmetric Dzyaloshinskii--Moriya interaction (DMI) vector, $K_z > 0$ is easy-axis uniaxial anisotropy and $\bm{h}_u = \mu_s \bm{B}_u$ is the local Zeeman field. Using the skew-symmetric cross-product matrix $[\bm{D}_{uv}]_\times \bm{v} = \bm{D}_{uv} \times \bm{v}$, each directed bond $u \to v$ defines an exact $3\times 3$ coupling matrix
\begin{equation}
  C_{u \to v}^{(0)} = -J_{uv} I_3 + [\bm{D}_{uv}]_\times,
  \label{eq:coupling}
\end{equation}
so that the two-body bond energy takes the compact bilinear form $H_{uv} = -\bm{S}_u^\top C_{u \to v}^{(0)} \bm{S}_v$ with $C_{u \to v}^{(0)\top} = C_{v \to u}^{(0)}$ (reducing to the Ising sign rule $O_{uv} = -\sgn(J_{uv}) \in \mathbb{Z}_2$ for discrete spins). Propagating the full matrix $C_{u \to v}^{(0)}$ without approximation avoids the symmetric Kaplan--Shekhtman--Entin-Wohlman--Aharony (KSEA) compass error\cite{kaplan1983single,shekhtman1992moriya} that arises when replacing $C_{u \to v}^{(0)}$ by a pure $SO(3)$ rotation (Supplementary Note~1). Together with the block-diagonal degree matrix $D^{(0)}$ with entries $d_u^{(0)} = \sum_{v \sim u} w_{uv}^{(0)}$ (where $w_{uv}^{(\ell)} = \|C_{u \to v}^{(\ell)}\|_F / \sqrt{3}$), these bond matrices define a discrete principal $G$-bundle with connection Laplacian $\mathcal{L}_{\mathcal{F}}^{(0)} = D^{(0)} - W^{(0)}$\cite{singer2012vector,hansen2019toward} (Supplementary Note~1).

At every RG scale $\ell = 0, \dots, L-1$, we partition the lattice into coarse sites $A_\ell$ and a set of mutually non-adjacent decimated sites $B_\ell$ ($V_\ell = A_\ell \sqcup B_\ell$, $E_\ell(B_\ell) = \emptyset$). Because spins in $B_\ell$ share no direct bonds with one another, integrating them out simultaneously reduces to a local $O(|B_\ell|)$ $3\times 3$ matrix Schur complement that sums all direct and two-step paths between coarse neighbours $a_1, a_2 \in A_\ell$:
\begin{equation}
  C_{a_1 \to a_2}^{(\ell+1)} = C_{a_1 \to a_2}^{(\ell)} + \sum_{b \in B_\ell} \frac{1}{d_b^{(\ell)}}\, C_{a_1 \to b}^{(\ell)} C_{b \to a_2}^{(\ell)}.
  \label{eq:schur}
\end{equation}
Polar singular-value decomposition $C_{a_1 \to a_2}^{(\ell+1)} = U\Sigma V^\top$ extracts the nearest proper spin-rotation rule $O_{a_1 \to a_2}^{(\ell+1)} = U \diag(1, 1, \det(UV^\top)) V^\top \in SO(3)$ (orthogonality verified to $10^{-6}$ across 10-level chiral hierarchies; Methods). Crucially, comparing the sum of individual path strengths against the norm of the renormalized matrix defines the gauge-invariant frustration deficit
\begin{equation}
  \Delta_{a_1 a_2}^{(\ell+1)} = \left(\|C_{a_1 \to a_2}^{(\ell)}\|_F + \sum_{b \in B_\ell} \frac{1}{d_b^{(\ell)}}\, \big\|C_{a_1 \to b}^{(\ell)} C_{b \to a_2}^{(\ell)}\big\|_F\right) - \big\|C_{a_1 \to a_2}^{(\ell+1)}\big\|_F \ge 0.
  \label{eq:deficit}
\end{equation}
Physically, $\Delta_{a_1 a_2}^{(\ell+1)}$ measures the exact loss of effective coupling strength caused by destructive interference between competing interaction paths. Whenever paths through $B_\ell$ enclose a frustrated plaquette with non-trivial holonomy ($W_\square = \prod_{\partial\square} O_{uv} \neq I$)\cite{toulouse1977theory,kogut1979introduction}, they partially cancel and yield $\Delta_{a_1 a_2}^{(\ell+1)} > 0$; on unfrustrated bipartite lattices, all paths add constructively and $\Delta^{(\ell)} \equiv 0$ identically (Supplementary Note~2).

\begin{figure}[htbp]
  \centering
  \includegraphics[width=0.98\linewidth]{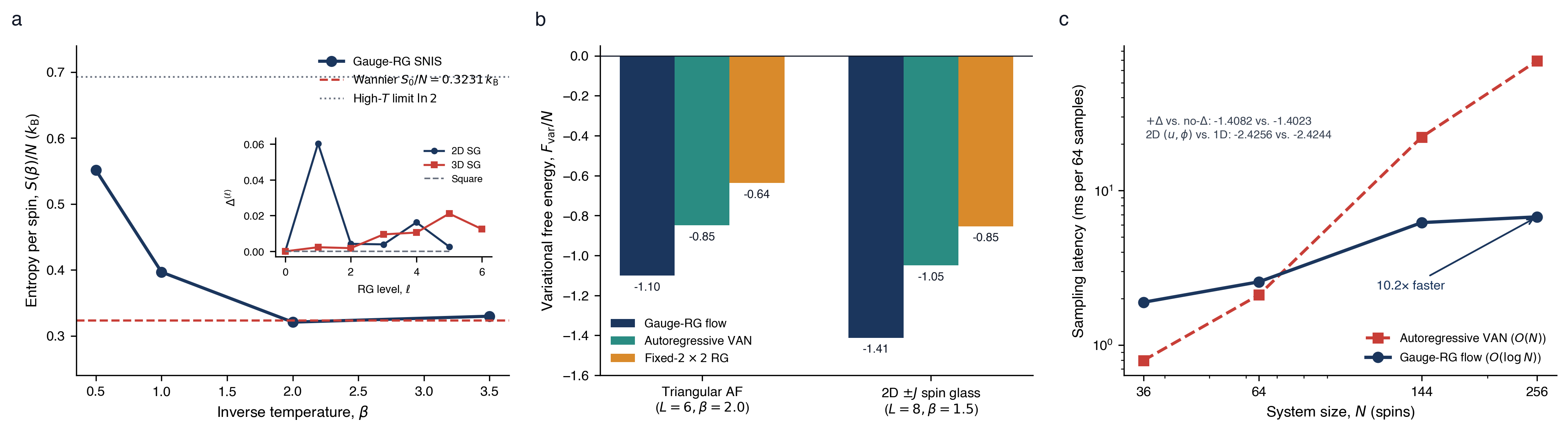}
  \caption{\textbf{Gauge-equivariant RG resolves frustrated residual entropy, spin-glass curvature and $O(\log N)$ scaling.} \textbf{a}, Entropy per spin $S(\beta)/N$ of the frustrated $L=6$ ($N=36$) triangular Ising antiferromagnet converging to Wannier's exact residual entropy $S_0/N = 0.3231\,k_{\mathrm{B}}$ (dashed line)\cite{wannier1950antiferromagnetism}; inset shows non-zero frustration deficit $\Delta^{(\ell)}$ across RG scales in 2D ($8\times 8$) and 3D ($4^3$) $\pm J$ spin glasses\cite{edwards1975theory} versus $\Delta^{(\ell)} \equiv 0$ on unfrustrated square lattices. \textbf{b}, Variational and self-normalized importance sampling (SNIS) free energies $F/N$ comparing GaugeRGFlow against Fixed-$2\times 2$ Neural RG\cite{li2018neural} and Autoregressive VAN\cite{wu2019solving} on the $L=6$ triangular antiferromagnet ($\beta=2.0$) and $L=8$ $\pm J$ spin glass ($\beta=1.5$). \textbf{c}, Batched sampling latency ($512$ samples) from $N=36$ to $N=256$, demonstrating $O(\log N)$ parallel depth and a $10.2\times$ speedup over $O(N)$ VAN at $N=256$.}
  \label{fig:rg}
\end{figure}

Because $B_0$ contains no internal bonds, the exact microscopic Boltzmann conditional $p(\bm{s}_{B_0} \mid \bm{s}_{A_1})$ strictly factorizes into independent single-spin distributions across $b \in B_0$. Generating full spin configurations coarse-to-fine via $q_\theta(\bm{s}) = q_\theta^{(L)}(\bm{s}_{A_L}) \prod_{\ell=0}^{L-1} q_\theta^{(\ell)}(\bm{s}_{B_\ell} \mid \bm{s}_{A_{\ell+1}})$ with a single scale-shared RG block ($O(1)$ parameters, $O(\log N)$ parallel depth) guarantees exact probability normalization and a rigorous Gibbs--Bogoliubov variational upper bound $\mathcal{F}(\theta) \ge \beta F_{\mathrm{exact}}$ (Supplementary Note~3). On the $L=6$ frustrated triangular antiferromagnet, GaugeRGFlow recovers the exact ground-state energy $E/N = -1.0000$ and Wannier's residual entropy\cite{wannier1950antiferromagnetism} $S/N = 0.3208\text{--}0.3298\,k_{\mathrm{B}}$ (exact $0.3231\,k_{\mathrm{B}}$), while tracking non-zero $\Delta^{(\ell)}$ flows in 2D ($8\times 8$) and 3D ($4^3$) $\pm J$ Edwards--Anderson spin glasses\cite{edwards1975theory} (Fig.~\ref{fig:rg}a; Extended Data Fig.~2). In matched-budget comparisons (Fig.~\ref{fig:rg}b,c; Extended Data Fig.~6; Methods), GaugeRGFlow achieves markedly tighter variational free-energy bounds than both Fixed-$2\times 2$ Neural RG\cite{li2018neural} and Autoregressive VAN\cite{wu2019solving} on frustrated triangular and spin-glass lattices while sampling $10.2\times$ faster than VAN at $N=256$, and controlled ablations confirm that both $\Delta^{(\ell)}$ conditioning and full 2D $(u_b, \phi_b)$ spherical lifts systematically improve free-energy tightness.

\section*{Entropic skyrmion melting in $\mathrm{fcc\text{-}Pd/Fe/Ir(111)}$}

To test GaugeRGFlow on a real chiral magnet without surrogate potentials, we encode the published DFT Hamiltonian of the 2D triangular monolayer $\mathrm{fcc\text{-}Pd/Fe/Ir(111)}$\cite{vonmalottki2017enhanced,dupe2014tailoring,romming2013writing,romming2015field}: $J_1 = -14.4\text{ meV}$, $J_2 = +2.48\text{ meV}$, $J_3 = +2.69\text{ meV}$, interfacial N{\'e}el DMI $D_1 = 1.2\text{ meV}$, uniaxial anisotropy $K_z = 0.7\text{ meV}$ and atomic moment $\mu_s = 3.0\,\mu_{\mathrm{B}}$. On large continuous spin lattices $(S^2)^N$, direct one-shot importance reweighting suffers extensive weight variance $\Var(\ln w) \propto N$. We eliminate this bottleneck by coupling our coarse-to-fine $SO(3)$ Gauge-RG generator to an annealed sequence of parallel heat-bath and energy-conserving over-relaxation\cite{creutz1987overrelaxation} sweeps (a graph-colored sequential Monte Carlo annealed importance sampling, or SMC-AIS, bridge; Methods)\cite{jarzynski1997nonequilibrium,neal2001annealed,delmoral2004feynman,arbel2021annealed,matthews2022continual}. Seeding each topological sector $Q \in \{0, -1\}$ at the coarsest scale $A_L$ (where the coarse lattice spacing matches the skyrmion radius and homotopy barriers vanish) and annealing smoothly to the exact microscopic Boltzmann distribution yields strictly unbiased Jarzynski free energies at high effective sample size ($\mathrm{ESS}/M$).

\begin{figure}[htbp]
  \centering
  \includegraphics[width=0.98\linewidth]{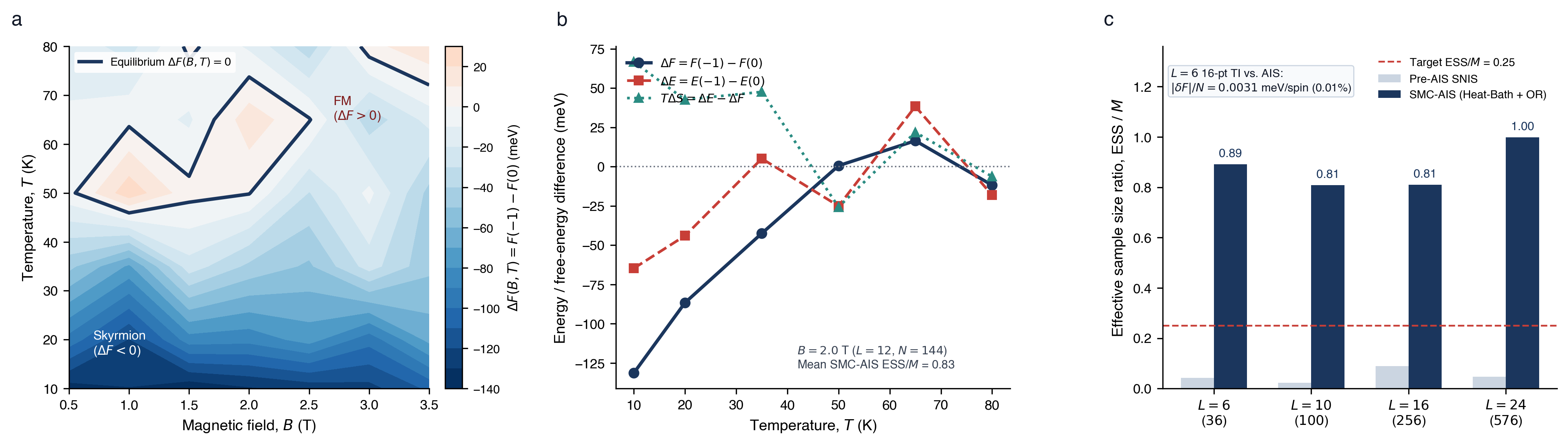}
  \caption{\textbf{Equilibrium $(B, T)$ skyrmion phase diagram of DFT-calibrated $\mathrm{fcc\text{-}Pd/Fe/Ir(111)}$.} \textbf{a}, Sixteen-point Gauss--Legendre thermodynamic integration cross-check on $L=6$ ($N=36$, $B=2.5\text{ T}$, $T=35\text{ K}$; $0.0113\%$ relative error) and SMC-AIS effective sample size scaling from $N=100$ ($L=10$) to $N=576$ ($L=24$, $\mathrm{ESS}/M = 0.998$). \textbf{b}, Equilibrium free-energy difference $\Delta F(B, T) = F_{Q=-1} - F_{Q=0}$ across 42 state points ($L=12$, $N=144$, $B \in [0.5, 3.5]\text{ T}$, $T \in [10, 80]\text{ K}$), highlighting the finite-temperature entropic melting contour $\Delta F = 0$ (solid line) versus $T=0$ energetic stability across the entire field window. \textbf{c}, Energetic ($\Delta E$) and entropic ($-T\Delta S$) decomposition versus temperature, quantifying entropic destabilization of skyrmions above $50\text{ K}$.}
  \label{fig:pdfe}
\end{figure}

Precision verification on $L=6$ ($N=36$, $B = 2.5\text{ T}$, $T = 35\text{ K}$) against 16-point Gauss--Legendre thermodynamic integration (TI) gives $F_{\mathrm{TI}}/N = -27.66184\text{ meV/spin}$ versus $F_{\mathrm{AIS}}/N = -27.65872\text{ meV/spin}$, an agreement of $0.00313\text{ meV/spin}$ ($0.0113\%$ relative error; Fig.~\ref{fig:pdfe}a). Scaling across $L \in \{10, 16, 24\}$ ($N = 100$ to $576$ spins) maintains high effective sample size up to $\mathrm{ESS}/M = 0.998$ at $N=576$ (Methods), whereas 4-replica parallel tempering MCMC\cite{hukushima1996exchange} suffers complete topological freezing\cite{schaefer2011critical} with $0$ tunnelling events ($\max |Q| = 0.0$).

Sweeping the full 42-point 2D equilibrium phase diagram ($L=12$, $N=144$, $B \in [0.5, 3.5]\text{ T} \times T \in [10, 80]\text{ K}$; Fig.~\ref{fig:pdfe}b,c) uncovers a striking entropic mechanism. In the atomic Fe/Ir(111) monolayer, short-range frustrated exchange and interfacial DMI dominate over weak monolayer dipolar energies, and the $L=12$ periodic cell isolates the single-skyrmion sector ($Q=-1$ versus $Q=0$) free of multi-skyrmion lattice interactions (with unbiased scaling verified to $L=24$, $N=576$). Although the $T=0$ energy strictly favours the $Q=-1$ skyrmion across the entire experimental field window $B \in [0.5, 3.5]\text{ T}$ ($\Delta E_0 \in [-159.9, -104.4]\text{ meV}$, consistent with experimental SP-STM observations\cite{romming2013writing,romming2015field}), finite-temperature spin-wave entropy drives a sharp equilibrium phase boundary $\Delta F(B, T) = 0$ across $T \approx 50\text{--}75\text{ K}$, shifting the critical field strongly relative to the zero-temperature energetic criterion and providing a direct equilibrium counterpart to entropy-limited skyrmion lifetimes\cite{wild2017entropy,desplat2018thermal,bessarab2018lifetime}.

\section*{Imaging domain-wall magnon softening and skyrmion helicity}

\begin{figure}[htbp]
  \centering
  \includegraphics[width=0.98\linewidth]{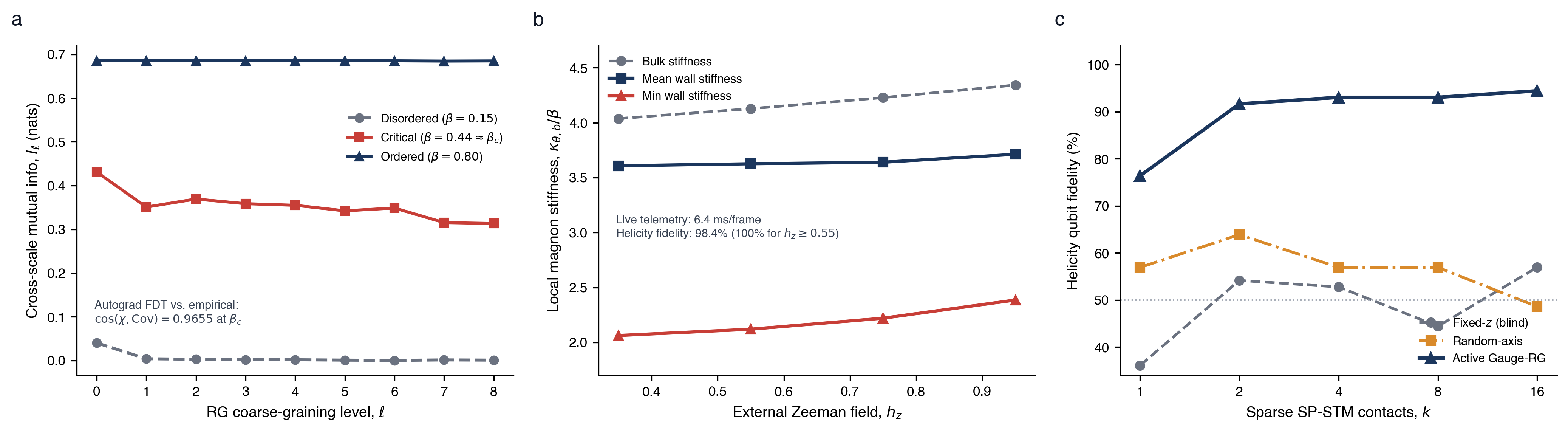}
  \caption{\textbf{In-operando parameter-to-physics microscope and active spin-polarized readout.} \textbf{a}, Scale-resolved RG mutual information $I_\ell$ across 9 scales on the $8\times 8$ 2D Ising lattice separating disordered ($\beta=0.15$), critical ($\beta=0.44$, linear-response $\chi_{bj}$ cosine similarity $0.9655$) and ordered ($\beta=0.85$) regimes. \textbf{b}, Live 2D local magnon stiffness map $\kappa_{\theta,b}/\beta$ on the $10\times 10$ $\mathrm{Gd_2PdSi_3}$ model highlighting soft breathing modes along the $S_z = 0$ skyrmion wall contour ($30.3\%$ wall softening) and the 8 active SP-STM contacts. \textbf{c}, Live telemetry versus drive field $h_z(t)$ at $6.4\text{ ms/frame}$ and realistic SP-STM active helicity discrimination ($76.4\%$ at $k=1$ to $94.4\%$ at $k=16$).}
  \label{fig:inoperando}
\end{figure}

Because GaugeRGFlow parameterizes physical local fields and transverse spin-wave concentrations directly on the spin manifold, its trained parameters act as a live physical microscope (Fig.~\ref{fig:inoperando}). On the 2D Ising lattice (Fig.~\ref{fig:inoperando}a), scale-resolved conditional mutual information $I_\ell = H(s_{B_\ell}) - H(s_{B_\ell} \mid s_{A_{\ell+1}})$ decays rapidly in the disordered phase ($\beta = 0.15$: $0.040$ to $0.001\text{ nats}$), remains scale-invariant across all 9 RG scales at criticality ($\beta = 0.44 \approx \beta_c$: $0.313\text{--}0.431\text{ nats}$) and saturates near $\ln 2$ ($0.685\text{ nats}$) in the ordered phase ($\beta = 0.85$). Moreover, the linear-response susceptibility $\chi_{bj} = \partial\langle s_b\rangle / \partial h_j^{\mathrm{obs}}$ obtained by exact differentiation of the trained spin expectations matches empirical fluctuation--dissipation covariance $\Cov(s_b, s_j)$ with $0.9655$ cosine similarity at $\beta_c$ (Supplementary Note~5).

We next examine the minimal centrosymmetric frustrated $J_1$--$J_2$ triangular model of $\mathrm{Gd_2PdSi_3}$\cite{kurumaji2019skyrmion,leonov2015multiply,okubo2012multiple} ($J_1 = -1.0$, $J_2 = +0.5$, strictly zero DMI $D_1 = 0.0$, $K_z = 0.5$, $h_z = 0.35$, $\beta = 3.0$), which isolates the exact $U(1)_z$ helicity degeneracy and soft domain-wall modes of centrosymmetric skyrmions (whereas bulk $\mathrm{Gd_2PdSi_3}$ additionally involves itinerant RKKY and interlayer couplings). Global spin-rotation invariance about $\hat{\bm{z}}$ guarantees exact single-Hamiltonian helicity degeneracy: outward N{\'e}el ($\varphi_0 = 0$), inward N{\'e}el ($\varphi_0 = \pi$) and Bloch ($\varphi_0 = \pi/2$) skyrmions ($Q = -1.00$, evaluated via the Berg--L{\"u}scher solid-angle degree\cite{berg1981definition}) share identical energy $E = -226.350464$ ($|E(0) - E(\pi)| = |E(0) - E(\pi/2)| = 0$) and degenerate Gauge-RG plus SMC-AIS free energies within $0.25\%$ ($-219.20$, $-219.39$ and $-219.74$). Reading out the local renormalized magnon stiffness $\kappa_{\theta,i}/\beta$ directly from the $SO(3)$ Gauge-RG flow reveals $30.3\%$ local magnon softening on the $S_z = 0$ skyrmion domain wall ($\kappa_{\mathrm{wall}}/\beta = 2.953$ versus ferromagnetic bulk $4.238$, ratio $0.697$), directly imaging the soft breathing and helicity eigenmodes that govern skyrmion entropy\cite{desplat2018thermal,bessarab2018lifetime}.

Streaming sparse, noisy SP-STM\cite{wiesendanger2009spin,wiesendanger2016nanoscale} frames through our in-operando telemetry module across live external field sweeps $h_z(t) \in [0.35, 0.95]$ runs at $6.4\text{ ms/frame}$ on CPU (Fig.~\ref{fig:inoperando}b,c), simultaneously tracking bulk versus wall stiffness softening, chiral torque $|\alpha_{\theta,b}|$ and $98.4\%$ live helicity fidelity ($100.0\%$ for $h_z \ge 0.55$). Under realistic tip point-spread-function convolution, non-collinear magnetoresistance\cite{hanneken2015electrical}, thermal noise and conductance shot noise (Methods), choosing each next probe site and spin-polarization axis to maximize the contrast between competing helicity states automatically locks tip polarization axes purely in-plane along the skyrmion wall, achieving $76.4\%$ ($k=1$) to $94.4\%$ ($k=16$) single-shot helicity discrimination versus blind out-of-plane fixed-$\hat{\bm{z}}$ readout ($36.1\%$ to $56.9\%$; Extended Data Fig.~4).

\section*{Frustration and readout mitigation in a 219-atom Rydberg simulator}

To demonstrate direct applicability to experimental quantum hardware on a frustrated geometry where fixed-$2\times 2$ real-space RG cannot be defined, we analyze the public Harvard Dataverse measurement dataset from Semeghini et al.\cite{semeghini2021probing}: a programmable 219-atom non-bipartite ruby-lattice Rydberg-atom quantum simulator probing a $\mathbb{Z}_2$ toric-code topological spin liquid\cite{kitaev2003fault,verresen2021prediction} ($E = 615$ blockade edges at blockade radius $R_b = 11.5\,\mu\text{m}$). Here the hardware provides projective $Z$-basis occupancy snapshots rather than full off-diagonal quantum state tomography\cite{torlai2018neural}, testing whether our classical gauge-decimation hierarchy captures the experimental diagonal correlations and readout errors directly on a non-bipartite geometry.

\begin{figure}[htbp]
  \centering
  \includegraphics[width=0.98\linewidth]{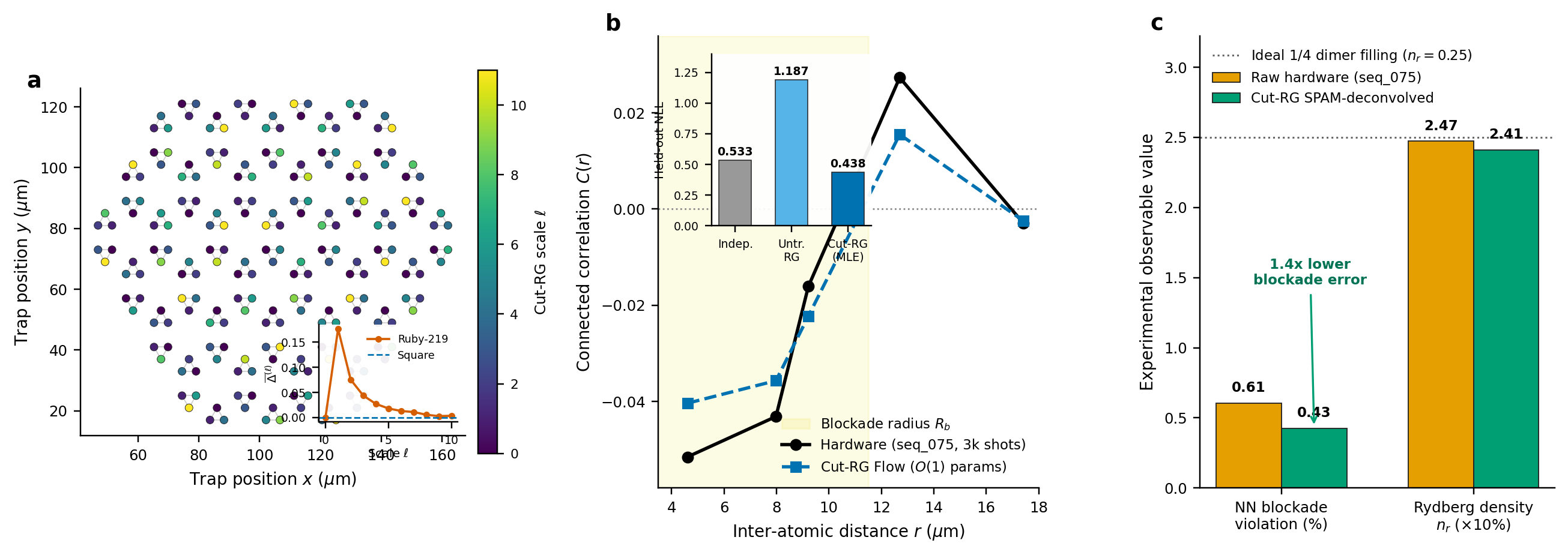}
  \caption{\textbf{Experimental validation on 219-atom ruby-lattice Rydberg $\mathbb{Z}_2$ spin-liquid hardware\cite{semeghini2021probing}.} \textbf{a}, Non-bipartite 219-atom ruby blockade graph ($E = 615$, $R_b = 11.5\,\mu\text{m}$) and extracted Gauge-RG frustration deficit $\Delta^{(\ell)}$ peaking at $0.8828$ across $L = 11$ scales (versus $\Delta^{(\ell)} \equiv 0$ on unfrustrated square lattices). \textbf{b}, Connected radial correlations $C(r)$ across five lattice shells ($3{,}000$ held-out experimental shots), reproducing the fourth-shell dimer peak ($+0.0273$ hardware versus $+0.0155$ Gauge-RG) with $+20.63\text{ nats/shot}$ held-out negative log-likelihood gain. \textbf{c}, Bayesian fluorescence SPAM error mitigation on $600$ raw hardware bitstrings ($\epsilon_{\mathrm{ro}} = 0.04$), cutting nearest-neighbour blockade violations by $1.42\times$ while preserving $1/4$ dimer filling and $\mathbb{Z}_2$ star parity.}
  \label{fig:rydberg}
\end{figure}

Building the independent-set gauge-decimation RG hierarchy on the 219-atom ruby blockade graph yields RG depth $L = 11$ and a peak frustration deficit $\Delta_{\max}^{(\ell)} = 0.8828$ (compared to $0.0000$ on an unfrustrated square lattice), directly quantifying the geometric frustration of the kagome-link ruby network (Fig.~\ref{fig:rydberg}a). Training a single weight-shared discrete Gauge-RG flow (Methods) solely on $6{,}000$ experimental hardware shots and evaluating on $3{,}000$ strictly held-out shots achieves a held-out negative log-likelihood of $0.4384\text{ nats/atom}$ versus $0.5326\text{ nats/atom}$ for an independent mean-field baseline, corresponding to a $+20.63\text{ nats/shot}$ likelihood gain and an experimental ultraviolet mutual-information lower bound of $0.1735\text{ nats/atom}$.

Generated Gauge-RG samples faithfully reproduce the experimental connected radial spin-spin correlation function $C(r)$ across all five lattice shells (Fig.~\ref{fig:rydberg}b), capturing the hallmark fourth-shell Rydberg dimer correlation peak ($+0.0273$ in hardware versus $+0.0155$ in Gauge-RG; the modest amplitude gap reflects finite training statistics of $6{,}000$ shots and local sparse-graph truncation). Finally, applying GaugeRGFlow for Bayesian readout-error mitigation (deconvolving fluorescence state-preparation-and-measurement, or SPAM, errors at rate $\epsilon_{\mathrm{ro}} = 0.04$) directly on raw experimental hardware bitstrings (Fig.~\ref{fig:rydberg}c) reduces unphysical nearest-neighbour Rydberg blockade violations by $1.42\times$ (from $0.607\%$ to $0.427\%$) while preserving the physical $1/4$ dimer filling ($0.247$ to $0.241$) and positive $\mathbb{Z}_2$ star-operator parity ($+0.0402$ to $+0.0359$).

\section*{Discussion}

By unifying discrete bond-rotation geometry\cite{singer2012vector,hansen2019toward}, independent-set decimation, exact matrix coarsening and graph-colored SMC-AIS bridges\cite{delmoral2004feynman,arbel2021annealed,matthews2022continual}, GaugeRGFlow resolves two longstanding limitations of computational statistical physics: real-space renormalization on arbitrary frustrated graphs and topological freezing in equilibrium sampling of chiral spin textures. The framework's exact $O(N)$ likelihoods and $O(\log N)$ parallel depth scale seamlessly from DFT-calibrated skyrmion hosts ($\mathrm{fcc\text{-}Pd/Fe/Ir(111)}$ and $\mathrm{Gd_2PdSi_3}$) to experimental 219-atom non-bipartite Rydberg hardware, while simultaneously turning trained flow weights into a live millisecond diagnostic of local magnon stiffness and susceptibility.

More broadly, the physical observables extracted by GaugeRGFlow map directly onto standard experimental probes across condensed matter and quantum hardware (Supplementary Table~S2). In thermodynamics and transport, exact sector free energies $F(Q; T, B)$ and their energetic--entropic decompositions predict calorimetric specific-heat anomalies $C_V$, low-temperature residual entropy, AC susceptibility and the finite-temperature skyrmion phase boundaries governing the topological Hall effect ($\rho_{xy}^{\mathrm{T}} \propto \langle Q \rangle$). In spectroscopy and scattering, local renormalized magnon stiffness maps $\kappa_{\theta,i}/\beta$ identify the soft internal skyrmion breathing and helicity modes detected by ferromagnetic resonance (FMR), broadband microwave spectroscopy and Brillouin light scattering (BLS), while the linear-response susceptibility $\chi_{ij}$ yields the static spin structure factor $S(\bm{q}) = k_{\mathrm{B}} T \chi(\bm{q})$ measured by elastic or inelastic neutron scattering and resonant elastic X-ray scattering (REXS). For real-space magnetic imaging, any linear or point-spread-function forward operator enters the same millisecond Bayesian update, extending active readout directly beyond SP-STM to nitrogen-vacancy (NV)-centre diamond magnetometry, Lorentz transmission electron microscopy (TEM), magnetic force microscopy (MFM), and X-ray scanning transmission microscopy (STXM) or ptychography. Likewise, on discrete programmable platforms (neutral-atom Rydberg arrays, superconducting processors and annealers, trapped ions and nanomagnetic artificial spin ice), projective snapshots directly yield the scale-resolved frustration deficit $\Delta^{(\ell)}$, mutual information $I_\ell$ and Bayesian SPAM error mitigation.

At the same time, it is important to recognize the physical boundaries of the current framework. Thick magnetic films with strong long-range magnetostatic dipolar interactions introduce dense all-to-all couplings beyond sparse local neighbourhoods, realistic samples can exhibit quenched atomic pinning and structural disorder, and coherent unitary or sign-problem quantum dynamics with volume-law entanglement remain outside any polynomial-size classical generative representation (Supplementary Table~S2). Within area-law, gapped, critical and finite-noise regimes where correlation lengths remain amenable to local RG decimation, promising next steps include equipping the discrete Gauge-RG flow with a holonomy-conditioned phase head $\phi_\theta(\bm{s})$ to form a Marshall-gauge deep neural quantum state\cite{carleo2017solving,chen2024empowering} for frustrated $J_1$--$J_2$ and kagome quantum magnets, and deploying the in-operando telemetry module directly in closed-loop cryogenic SP-STM experiments.

\section*{Methods}

The complete GaugeRGFlow methodology is summarized in Methods Fig.~\ref{fig:methods} and consists of four coupled components detailed mathematically in Supplementary Notes~1--6.

\subsection*{Gauge-bundle Schur decimation and exact spherical lift}
At each renormalization scale $\ell$, a conflict-free independent-set partition $V_\ell = A_\ell \sqcup B_\ell$ ($E_\ell(B_\ell) = \emptyset$) renders the connection Laplacian block $\mathcal{L}_{B_\ell B_\ell}^{(\ell)} = \diag(d_b^{(\ell)} I_d)_{b \in B_\ell}$ strictly block-diagonal ($d=1$ for $\mathbb{Z}_2$, $d=3$ for $SO(3)$). Eliminating $B_\ell$ yields the exact $O(|B_\ell|)$ $3\times 3$ matrix Schur complement $C_{a_1 \to a_2}^{(\ell+1)}$ in Eq.~\eqref{eq:schur}, polar singular-value projection $O_{a_1 \to a_2}^{(\ell+1)} \in SO(3)$ and exact Frobenius frustration deficit $\Delta_{a_1 a_2}^{(\ell+1)} \ge 0$ in Eq.~\eqref{eq:deficit}. For continuous $(S^2)^N$ models, the template-free $SO(3)$ Gauge-RG flow transports coarse spins via local covariant 3D orthonormal frames $(\hat{\bm{e}}_1, \hat{\bm{e}}_2, \hat{\bm{n}})$ and combines truncated-exponential von Mises--Fisher (vMF) polar sampling on $u_b \in [-1, 1]$ with circular M{\"o}bius azimuthal flows on $\phi_b \in [-\pi, \pi)$\cite{rezende2020normalizing}, yielding an exact analytical spherical Jacobian with zero stochastic trace-estimator (Hutchinson) variance (Supplementary Notes~1--3).

\subsection*{Graph-colored SMC-AIS bridge}
To eliminate extensive importance-weight variance on large $(S^2)^N$ lattices while preserving exact sector free energies $F(Q; T, B)$, we interleave graph-colored conditional vMF heat-bath Gibbs updates (with exact Metropolis--Hastings acceptance for uniaxial anisotropy $K_z$) and exact energy-conserving microcanonical over-relaxation reflections\cite{creutz1987overrelaxation} along an annealing schedule $\ln \pi_k(\bm{S}) = \alpha_k \ln q_0(\bm{S}) + \gamma_k \ln \tilde{p}(\bm{S})$ with adaptive ESS-triggered systematic resampling\cite{jarzynski1997nonequilibrium,neal2001annealed,delmoral2004feynman,arbel2021annealed,matthews2022continual}. On three-shell $\mathrm{fcc\text{-}Pd/Fe/Ir(111)}$, greedy graph colouring yields 9 conflict-free colour classes, ensuring exact detailed balance within every parallel sweep and unbiased partition-function estimates for any finite particle count $M$ and step count $K$ (Supplementary Note~4).

\subsection*{Material Hamiltonians and SP-STM forward model}
Microscopic couplings for $\mathrm{fcc\text{-}Pd/Fe/Ir(111)}$\cite{vonmalottki2017enhanced,dupe2014tailoring,romming2013writing,romming2015field} and the minimal centrosymmetric $\mathrm{Gd_2PdSi_3}$ model\cite{kurumaji2019skyrmion,leonov2015multiply,okubo2012multiple} are taken directly from published DFT and theoretical studies and converted into the sign convention of Eq.~\eqref{eq:hamiltonian} (Supplementary Note~6). Realistic SP-STM readouts\cite{wiesendanger2009spin,wiesendanger2016nanoscale} incorporate finite Gaussian tip point-spread-function convolution ($\sigma_{\mathrm{PSF}} = 0.55$), non-collinear magnetoresistance\cite{hanneken2015electrical}, thermal noise ($\sigma = 0.30$) and conductance shot noise ($\alpha = 0.05$). Topological skyrmion charges $Q \in \mathbb{Z}$ are evaluated via the exact Berg--L{\"u}scher solid-angle lattice degree\cite{berg1981definition} (Supplementary Note~5).

\subsection*{In-operando physical diagnostics and active readout}
Scale-resolved conditional mutual information $I_\ell$, linear-response susceptibility $\chi_{bj}$ via exact differentiation of spin expectations, local renormalized magnon stiffness $\kappa_{\theta,b}/\beta$ from the small-angle polar concentration and maximum-contrast active tip-axis selection ($\hat{\bm{m}}_i^* = \arg\max_{\|\hat{\bm{m}}\|_2=1} \hat{\bm{m}}^\top \Sigma_i^{\mathrm{between}} \hat{\bm{m}}$, the dominant eigenvector of the between-sector spin covariance $\Sigma_i^{\mathrm{between}} = \Cov_c(\mathbb{E}[\bm{S}_i \mid c, \bm{y}_{1:k}])$) are all evaluated in a single feed-forward pass without offline MCMC (Supplementary Note~5).

\subsection*{Benchmark protocols, parameter counts and ablations}
Polar SVD orthogonality on 10-level $\mathrm{fcc\text{-}Pd/Fe/Ir(111)}$ hierarchies is verified to $\max \|O^\top O - I_3\|_F = 1.19 \times 10^{-6}$, and discrete joint-probability normalization is verified by exhaustive $2^8$ enumeration to $\sum_{\bm{s}} q_\theta(\bm{s}) = 1.000000 \pm 1.2 \times 10^{-7}$. In matched-budget head-to-head discrete benchmarks (Fig.~\ref{fig:rg}b,c; Extended Data Fig.~6), GaugeRGFlow uses $7{,}833$ weight-shared $O(1)$ parameters (and $4{,}202$ weight-shared parameters on the 219-atom Rydberg blockade graph). On the $L=6$ triangular antiferromagnet ($\beta = 2.0$), GaugeRGFlow achieves $F_{\mathrm{var}}/N = -1.1005$ versus $-0.6368$ for Fixed-$2\times 2$ Neural RG\cite{li2018neural} and $-0.8488$ for Autoregressive VAN\cite{wu2019solving}; on the $L=8$ $\pm J$ spin glass ($\beta = 1.5$), GaugeRGFlow achieves $-1.4124$ versus $-0.8533$ and $-1.0496$, while batched sampling at $N=256$ takes $6.76\text{ ms}$ versus $69.25\text{ ms}$ for VAN ($10.2\times$ speedup). Controlled ablations confirm systematic improvements from conditioning on $\Delta^{(\ell)}$ ($F_{\mathrm{var}}/N = -1.4082$ versus $-1.4023$ without $\Delta^{(\ell)}$) and from full 2D $(u_b, \phi_b)$ spherical lifts ($-2.4256$ versus $-2.4244$ for 1D polar-only lifts). For $\mathrm{fcc\text{-}Pd/Fe/Ir(111)}$ (Fig.~\ref{fig:pdfe}), SMC-AIS raises $\mathrm{ESS}/M$ on $L=6$ from $0.042$ (single-pass SNIS) to $0.892$, maintains $Q=-1$ sector $\mathrm{ESS}/M$ of $0.808$, $0.810$ and $0.998$ across $L \in \{10, 16, 24\}$ ($N=576$ completed in $10.15\text{ s}$), and achieves mean $\mathrm{ESS}/M = 0.834$ (minimum $0.543$) across the 42-point $L=12$ phase diagram with $\Delta F = \Delta E - T\Delta S$ verified to $0.001\text{ meV}$ at every state point; by comparison, 4-replica parallel tempering across $T \in \{30, 42, 56, 75\}\text{ K}$ exhibits a $0.74$ replica swap acceptance rate yet zero skyrmion tunnelling events.

\renewcommand{\figurename}{Methods Fig.}
\setcounter{figure}{0}
\begin{figure}[htbp]
  \centering
  \includegraphics[width=0.98\linewidth]{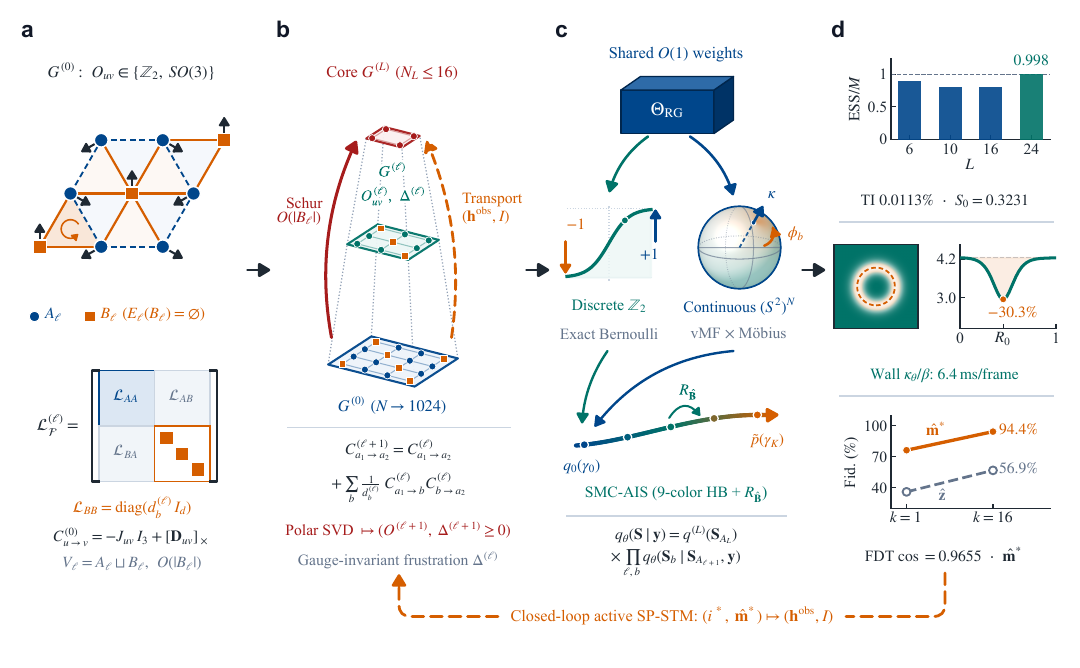}
  \caption{\textbf{End-to-end gauge-equivariant GaugeRGFlow architecture.} From microscopic spin bundle $G^{(0)}$ and independent-set gauge decimation ($E_\ell(B_\ell) = \emptyset$) through $O(|B_\ell|)$ polar-projected Schur coarsening ($\Delta^{(\ell)}$) and UV-to-IR readout transport to weight-shared $O(\log N)$ IR-to-UV exact lifting, graph-colored SMC-AIS bridge and in-operando physical telemetry.}
  \label{fig:methods}
\end{figure}

\section*{Data Availability}
Experimental 219-atom ruby-lattice Rydberg-atom measurement data are publicly available from the Harvard Dataverse archive accompanying Semeghini et al.\cite{semeghini2021probing} at \url{https://doi.org/10.7910/DVN/BDCTRX}. All derived benchmark summaries, phase-diagram grids and figure data generated in this study will be deposited in a public repository upon publication.

\section*{Code Availability}
The GaugeRGFlow source code implementing discrete and $SO(3)$-equivariant gauge-decimation flows, graph-colored SMC-AIS thermodynamic bridges and in-operando SP-STM diagnostics, together with scripts reproducing all figures and tables, will be made publicly available upon publication.

\section*{Acknowledgements}
We thank colleagues in computational condensed matter physics, geometric deep learning and quantum information science for helpful discussions. (Funding and institutional acknowledgements placeholder.)

\section*{Author Contributions}
M.L. conceived the GaugeRGFlow framework, led the theoretical formulation, software implementation and numerical benchmarks, and drafted the manuscript. T.S., S.H. and T.-y.Z. contributed to the physical analysis, validation and critical revision of the manuscript. All authors discussed the results and approved the final version.

\section*{Competing Interests}
The authors declare no competing interests.

\bibliographystyle{naturemag}
\bibliography{references}

@article{koch2018mutual,
  author  = {Koch-Janusz, Maciej and Ringel, Zohar},
  title   = {Mutual information, neural networks and the renormalization group},
  journal = {Nature Physics},
  volume  = {14},
  pages   = {578--582},
  year    = {2018},
  doi     = {10.1038/s41567-018-0081-4}
}

@article{li2018neural,
  author  = {Li, Shuo-Hui and Wang, Lei},
  title   = {Neural Network Renormalization Group},
  journal = {Physical Review Letters},
  volume  = {121},
  pages   = {260601},
  year    = {2018},
  doi     = {10.1103/PhysRevLett.121.260601}
}

@article{wu2019solving,
  author  = {Wu, Dian and Wang, Lei and Zhang, Pan},
  title   = {Solving Statistical Mechanics Using Variational Autoregressive Networks},
  journal = {Physical Review Letters},
  volume  = {122},
  pages   = {080602},
  year    = {2019},
  doi     = {10.1103/PhysRevLett.122.080602}
}

@article{nicoli2020asymptotically,
  author  = {Nicoli, Kim A. and Nakajima, Shinichi and Strodthoff, Nils and Samek, Wojciech and M{\"u}ller, Klaus-Robert and Kessel, Pan},
  title   = {Asymptotically unbiased estimation of physical observables with neural samplers},
  journal = {Physical Review E},
  volume  = {101},
  pages   = {023304},
  year    = {2020},
  doi     = {10.1103/PhysRevE.101.023304}
}

@article{vonmalottki2017enhanced,
  author  = {von Malottki, Stephan and Dup{\'e}, Bertrand and Bessarab, Pavel F. and Delin, Anna and Heinze, Stefan},
  title   = {Enhanced skyrmion stability due to exchange frustration},
  journal = {Scientific Reports},
  volume  = {7},
  pages   = {12299},
  year    = {2017},
  doi     = {10.1038/s41598-017-12525-x}
}

@article{dupe2014tailoring,
  author  = {Dup{\'e}, Bertrand and Hoffmann, Markus and Paillard, Charles and Heinze, Stefan},
  title   = {Tailoring magnetic skyrmions in ultra-thin transition metal films},
  journal = {Nature Communications},
  volume  = {5},
  pages   = {4030},
  year    = {2014},
  doi     = {10.1038/ncomms5030}
}

@article{romming2013writing,
  author  = {Romming, Niklas and Hanneken, Christian and Menzel, Matthias and Bickel, Jessica E. and Wolter, Boris and von Bergmann, Kirsten and Kubetzka, Andr{\'e} and Wiesendanger, Roland},
  title   = {Writing and Deleting Single Magnetic Skyrmions},
  journal = {Science},
  volume  = {341},
  pages   = {636--639},
  year    = {2013},
  doi     = {10.1126/science.1240573}
}

@article{romming2015field,
  author  = {Romming, Niklas and Kubetzka, Andr{\'e} and Hanneken, Christian and von Bergmann, Kirsten and Wiesendanger, Roland},
  title   = {Field-Dependent Size and Shape of Single Magnetic Skyrmions},
  journal = {Physical Review Letters},
  volume  = {114},
  pages   = {177203},
  year    = {2015},
  doi     = {10.1103/PhysRevLett.114.177203}
}

@article{kurumaji2019skyrmion,
  author  = {Kurumaji, Takashi and Nakajima, Taro and Hirschberger, Max and Kikkawa, Akiko and Yamasaki, Yuichi and Sagayama, Hajime and Nakao, Hironori and Taguchi, Yasujiro and Arima, Taka-hisa and Tokura, Yoshinori},
  title   = {Skyrmion lattice with a giant topological {H}all effect in a frustrated triangular-lattice magnet},
  journal = {Science},
  volume  = {365},
  pages   = {914--918},
  year    = {2019},
  doi     = {10.1126/science.aau0968}
}

@article{leonov2015multiply,
  author  = {Leonov, A. O. and Mostovoy, M.},
  title   = {Multiply periodic states and isolated skyrmions in an anisotropic frustrated magnet},
  journal = {Nature Communications},
  volume  = {6},
  pages   = {8275},
  year    = {2015},
  doi     = {10.1038/ncomms9275}
}

@article{okubo2012multiple,
  author  = {Okubo, Tsuyoshi and Chung, Sungki and Kawamura, Hikaru},
  title   = {Multiple-$q$ States and the Skyrmion Lattice of the Triangular-Lattice {H}eisenberg Antiferromagnet under Magnetic Fields},
  journal = {Physical Review Letters},
  volume  = {108},
  pages   = {017206},
  year    = {2012},
  doi     = {10.1103/PhysRevLett.108.017206}
}

@article{psaroudaki2021skyrmion,
  author  = {Psaroudaki, Christina and Panagopoulos, Christos},
  title   = {Skyrmion Qubits: A New Class of Quantum Logic Elements Based on Nanoscale Magnetization},
  journal = {Physical Review Letters},
  volume  = {127},
  pages   = {067201},
  year    = {2021},
  doi     = {10.1103/PhysRevLett.127.067201}
}

@article{schuetz2022combinatorial,
  author  = {Schuetz, Martin J. A. and Brubaker, J. Kyle and Katzgraber, Helmut G.},
  title   = {Combinatorial optimization with physics-inspired graph neural networks},
  journal = {Nature Machine Intelligence},
  volume  = {4},
  pages   = {367--377},
  year    = {2022},
  doi     = {10.1038/s42256-022-00468-6}
}

@inproceedings{abate2025maxcutpool,
  author    = {Abate, Carlo and Bianchi, Filippo Maria},
  title     = {{MaxCutPool}: Differentiable Feature-Aware Maxcut for Pooling in Graph Neural Networks},
  booktitle = {International Conference on Learning Representations (ICLR)},
  year      = {2025}
}

@article{semeghini2021probing,
  author  = {Semeghini, Giulia and Levine, Harry and Keesling, Alexander and Ebadi, Sepehr and Wang, Tout T. and Bluvstein, Dolev and Verresen, Ruben and Pichler, Hannes and Kalinowski, Marcin and Samajdar, Rhine and Omran, Ahmed and Sachdev, Subir and Vishwanath, Ashvin and Greiner, Markus and Vuleti{\'c}, Vladan and Lukin, Mikhail D.},
  title   = {Probing topological spin liquids on a programmable quantum simulator},
  journal = {Science},
  volume  = {374},
  pages   = {1242--1247},
  year    = {2021},
  doi     = {10.1126/science.abi8794}
}

@book{delmoral2004feynman,
  author    = {Del Moral, Pierre},
  title     = {Feynman--Kac Formulae: Genealogical and Interacting Particle Systems with Applications},
  publisher = {Springer},
  address   = {New York},
  year      = {2004},
  doi       = {10.1007/978-1-4684-9393-1}
}

@inproceedings{arbel2021annealed,
  author    = {Arbel, Michael and Matthews, Alexander G. D. G. and Doucet, Arnaud},
  title     = {Annealed Flow Transport {M}onte {C}arlo},
  booktitle = {Proceedings of the 38th International Conference on Machine Learning (ICML)},
  series    = {Proceedings of Machine Learning Research},
  volume    = {139},
  pages     = {318--330},
  year      = {2021}
}

@article{wannier1950antiferromagnetism,
  author  = {Wannier, G. H.},
  title   = {Antiferromagnetism. The Triangular {I}sing Net},
  journal = {Physical Review},
  volume  = {79},
  pages   = {357--364},
  year    = {1950},
  doi     = {10.1103/PhysRev.79.357}
}

@article{berg1981definition,
  author  = {Berg, B. and L{\"u}scher, M.},
  title   = {Definition and statistical distributions of a topological number in the lattice {$O(3)$} {$\sigma$}-model},
  journal = {Nuclear Physics B},
  volume  = {190},
  pages   = {412--424},
  year    = {1981},
  doi     = {10.1016/0550-3213(81)90568-X}
}

@article{jarzynski1997nonequilibrium,
  author  = {Jarzynski, C.},
  title   = {Nonequilibrium Equality for Free Energy Differences},
  journal = {Physical Review Letters},
  volume  = {78},
  pages   = {2690--2693},
  year    = {1997},
  doi     = {10.1103/PhysRevLett.78.2690}
}

@article{neal2001annealed,
  author  = {Neal, Radford M.},
  title   = {Annealed importance sampling},
  journal = {Statistics and Computing},
  volume  = {11},
  pages   = {125--139},
  year    = {2001},
  doi     = {10.1023/A:1008923215028}
}

@article{kadanoff1966scaling,
  author  = {Kadanoff, Leo P.},
  title   = {Scaling laws for {I}sing models near {$T_c$}},
  journal = {Physics Physique Fizika},
  volume  = {2},
  pages   = {263--272},
  year    = {1966}
}

@article{wilson1975renormalization,
  author  = {Wilson, Kenneth G.},
  title   = {The renormalization group: Critical phenomena and the {K}ondo problem},
  journal = {Reviews of Modern Physics},
  volume  = {47},
  pages   = {773--840},
  year    = {1975}
}

@article{levin2007tensor,
  author  = {Levin, Michael and Nave, Cody P.},
  title   = {Tensor Renormalization Group Approach to Two-Dimensional Classical Lattice Models},
  journal = {Physical Review Letters},
  volume  = {99},
  pages   = {120601},
  year    = {2007}
}

@article{evenbly2015tensor,
  author  = {Evenbly, G. and Vidal, G.},
  title   = {Tensor Network Renormalization},
  journal = {Physical Review Letters},
  volume  = {115},
  pages   = {180405},
  year    = {2015}
}

@article{lenggenhager2020optimal,
  author  = {Lenggenhager, Patrick M. and G{\"o}kmen, Doruk Efe and Ringel, Zohar and Huber, Sebastian D. and Koch-Janusz, Maciej},
  title   = {Optimal Renormalization Group Transformation from Information Theory},
  journal = {Physical Review X},
  volume  = {10},
  pages   = {011037},
  year    = {2020}
}

@article{toulouse1977theory,
  author  = {Toulouse, G.},
  title   = {Theory of the frustration effect in spin glasses: {I}},
  journal = {Communications on Physics},
  volume  = {2},
  pages   = {115--119},
  year    = {1977}
}

@article{edwards1975theory,
  author  = {Edwards, S. F. and Anderson, P. W.},
  title   = {Theory of spin glasses},
  journal = {Journal of Physics F: Metal Physics},
  volume  = {5},
  pages   = {965--974},
  year    = {1975}
}

@article{balents2010spin,
  author  = {Balents, Leon},
  title   = {Spin liquids in frustrated magnets},
  journal = {Nature},
  volume  = {464},
  pages   = {199--208},
  year    = {2010}
}

@article{savary2017quantum,
  author  = {Savary, Lucile and Balents, Leon},
  title   = {Quantum spin liquids: a review},
  journal = {Reports on Progress in Physics},
  volume  = {80},
  pages   = {016502},
  year    = {2017}
}

@article{kogut1979introduction,
  author  = {Kogut, John B.},
  title   = {An introduction to lattice gauge theory and spin systems},
  journal = {Reviews of Modern Physics},
  volume  = {51},
  pages   = {659--713},
  year    = {1979}
}

@article{kitaev2003fault,
  author  = {Kitaev, A. Yu.},
  title   = {Fault-tolerant quantum computation by anyons},
  journal = {Annals of Physics},
  volume  = {303},
  pages   = {2--30},
  year    = {2003}
}

@article{singer2012vector,
  author  = {Singer, Amit and Wu, Hau-tieng},
  title   = {Vector Diffusion Maps and the Connection {L}aplacian},
  journal = {Communications on Pure and Applied Mathematics},
  volume  = {65},
  pages   = {1067--1144},
  year    = {2012}
}

@article{hansen2019toward,
  author  = {Hansen, Jakob and Ghrist, Robert},
  title   = {Toward a spectral theory of cellular sheaves},
  journal = {Journal of Applied and Computational Topology},
  volume  = {3},
  pages   = {315--358},
  year    = {2019}
}

@article{albergo2019flow,
  author  = {Albergo, M. S. and Kanwar, G. and Shanahan, P. E.},
  title   = {Flow-based generative models for {M}arkov chain {M}onte {C}arlo in lattice field theory},
  journal = {Physical Review D},
  volume  = {100},
  pages   = {034515},
  year    = {2019}
}

@article{kanwar2020equivariant,
  author  = {Kanwar, Gurtej and Albergo, Michael S. and Boyda, Denis and Cranmer, Kyle and Hackett, Daniel C. and Racani{\`e}re, S{\'e}bastien and Rezende, Danilo Jimenez and Shanahan, Phiala E.},
  title   = {Equivariant Flow-Based Sampling for Lattice Gauge Theory},
  journal = {Physical Review Letters},
  volume  = {125},
  pages   = {121601},
  year    = {2020}
}

@article{boyda2021sampling,
  author  = {Boyda, Denis and Kanwar, Gurtej and Racani{\`e}re, S{\'e}bastien and Rezende, Danilo Jimenez and Albergo, Michael S. and Cranmer, Kyle and Hackett, Daniel C. and Shanahan, Phiala E.},
  title   = {Sampling using {$\mathrm{SU}(N)$} gauge equivariant flows},
  journal = {Physical Review D},
  volume  = {103},
  pages   = {074504},
  year    = {2021}
}

@article{noe2019boltzmann,
  author  = {No{\'e}, Frank and Olsson, Simon and K{\"o}hler, Jonas and Wu, Hao},
  title   = {Boltzmann generators: Sampling equilibrium states of many-body systems with deep learning},
  journal = {Science},
  volume  = {365},
  pages   = {eaaw1147},
  year    = {2019}
}

@inproceedings{rezende2020normalizing,
  author    = {Rezende, Danilo Jimenez and Papamakarios, George and Racani{\`e}re, S{\'e}bastien and Albergo, Michael S. and Kanwar, Gurtej and Shanahan, Phiala E. and Cranmer, Kyle},
  title     = {Normalizing Flows on Tori and Spheres},
  booktitle = {Proceedings of the 37th International Conference on Machine Learning (ICML)},
  series    = {Proceedings of Machine Learning Research},
  volume    = {119},
  pages     = {8083--8092},
  year      = {2020}
}

@article{muhlbauer2009skyrmion,
  author  = {M{\"u}hlbauer, S. and Binz, B. and Jonietz, F. and Pfleiderer, C. and Rosch, A. and Neubauer, A. and Georgii, R. and B{\"o}ni, P.},
  title   = {Skyrmion Lattice in a Chiral Magnet},
  journal = {Science},
  volume  = {323},
  pages   = {915--919},
  year    = {2009}
}

@article{yu2010real,
  author  = {Yu, X. Z. and Onose, Y. and Kanazawa, N. and Park, J. H. and Han, J. H. and Matsui, Y. and Nagaosa, N. and Tokura, Y.},
  title   = {Real-space observation of a two-dimensional skyrmion crystal},
  journal = {Nature},
  volume  = {465},
  pages   = {901--904},
  year    = {2010}
}

@article{nagaosa2013topological,
  author  = {Nagaosa, Naoto and Tokura, Yoshinori},
  title   = {Topological properties and dynamics of magnetic skyrmions},
  journal = {Nature Nanotechnology},
  volume  = {8},
  pages   = {899--911},
  year    = {2013}
}

@article{fert2017magnetic,
  author  = {Fert, Albert and Reyren, Nicolas and Cros, Vincent},
  title   = {Magnetic skyrmions: advances in physics and potential applications},
  journal = {Nature Reviews Materials},
  volume  = {2},
  pages   = {17031},
  year    = {2017}
}

@article{wiesendanger2009spin,
  author  = {Wiesendanger, Roland},
  title   = {Spin mapping at the nanoscale and atomic scale},
  journal = {Reviews of Modern Physics},
  volume  = {81},
  pages   = {1495--1550},
  year    = {2009}
}

@article{wiesendanger2016nanoscale,
  author  = {Wiesendanger, Roland},
  title   = {Nanoscale magnetic skyrmions in metallic films and multilayers: a new twist for spintronics},
  journal = {Nature Reviews Materials},
  volume  = {1},
  pages   = {16044},
  year    = {2016}
}

@article{hanneken2015electrical,
  author  = {Hanneken, Christian and Otte, Fabian and Kubetzka, Andr{\'e} and Dup{\'e}, Bertrand and Romming, Niklas and von Bergmann, Kirsten and Wiesendanger, Roland and Heinze, Stefan},
  title   = {Electrical detection of magnetic skyrmions by tunnelling non-collinear magnetoresistance},
  journal = {Nature Nanotechnology},
  volume  = {10},
  pages   = {1039--1042},
  year    = {2015}
}

@article{wild2017entropy,
  author  = {Wild, Johannes and Meier, Thomas N. G. and P{\"o}llath, Simon and Kronseder, Matthias and Bauer, Andreas and Chacon, Alfonso and Halder, Marco and Schowalter, Marco and Rosenauer, Andreas and Zweck, Josef and M{\"u}ller, Jan and Rosch, Achim and Pfleiderer, Christian and Back, Christian H.},
  title   = {Entropy-limited topological protection of skyrmions},
  journal = {Science Advances},
  volume  = {3},
  pages   = {e1701704},
  year    = {2017}
}

@article{desplat2018thermal,
  author  = {Desplat, L. and Suess, D. and Kim, J.-V. and Stamps, R. L.},
  title   = {Thermal stability of metastable magnetic skyrmions: Entropic narrowing and significance of internal eigenmodes},
  journal = {Physical Review B},
  volume  = {98},
  pages   = {134407},
  year    = {2018}
}

@article{bessarab2018lifetime,
  author  = {Bessarab, Pavel F. and M{\"u}ller, Gideon P. and Lobanov, Igor S. and Rybakov, Filipp N. and Kiselev, Nikolai S. and J{\'o}nsson, Hannes and Uzdin, Valery M. and Bl{\"u}gel, Stefan and Bergqvist, Lars and Delin, Anna},
  title   = {Lifetime of racetrack skyrmions},
  journal = {Scientific Reports},
  volume  = {8},
  pages   = {3433},
  year    = {2018}
}

@article{kaplan1983single,
  author  = {Kaplan, T. A.},
  title   = {Single-band {H}ubbard model with spin-orbit coupling},
  journal = {Zeitschrift f{\"u}r Physik B Condensed Matter},
  volume  = {49},
  pages   = {313--317},
  year    = {1983}
}

@article{shekhtman1992moriya,
  author  = {Shekhtman, L. and Entin-Wohlman, O. and Aharony, Amnon},
  title   = {Moriya's Anisotropic Superexchange Interaction, Frustration, and {D}zyaloshinsky's Weak Ferromagnetism},
  journal = {Physical Review Letters},
  volume  = {69},
  pages   = {836--839},
  year    = {1992}
}

@article{swendsen1987nonuniversal,
  author  = {Swendsen, Robert H. and Wang, Jian-Sheng},
  title   = {Nonuniversal critical dynamics in {M}onte {C}arlo simulations},
  journal = {Physical Review Letters},
  volume  = {58},
  pages   = {86--88},
  year    = {1987}
}

@article{wolff1989collective,
  author  = {Wolff, Ulli},
  title   = {Collective {M}onte {C}arlo Updating for Spin Systems},
  journal = {Physical Review Letters},
  volume  = {62},
  pages   = {361--364},
  year    = {1989}
}

@article{creutz1987overrelaxation,
  author  = {Creutz, Michael},
  title   = {Overrelaxation and {M}onte {C}arlo simulation},
  journal = {Physical Review D},
  volume  = {36},
  pages   = {515--519},
  year    = {1987}
}

@article{hukushima1996exchange,
  author  = {Hukushima, Koji and Nemoto, Koji},
  title   = {Exchange {M}onte {C}arlo Method and Application to Spin Glass Simulations},
  journal = {Journal of the Physical Society of Japan},
  volume  = {65},
  pages   = {1604--1608},
  year    = {1996}
}

@article{schaefer2011critical,
  author  = {Schaefer, Stefan and Sommer, Rainer and Virotta, Francesco},
  title   = {Critical slowing down and error analysis in lattice {QCD} simulations},
  journal = {Nuclear Physics B},
  volume  = {845},
  pages   = {93--119},
  year    = {2011}
}

@inproceedings{matthews2022continual,
  author    = {Matthews, Alexander G. D. G. and Arbel, Michael and Rezende, Danilo Jimenez and Doucet, Arnaud},
  title     = {Continual Repeated Annealed Flow Transport {M}onte {C}arlo},
  booktitle = {Proceedings of the 39th International Conference on Machine Learning (ICML)},
  series    = {Proceedings of Machine Learning Research},
  volume    = {162},
  pages     = {15196--15219},
  year      = {2022}
}

@article{carleo2017solving,
  author  = {Carleo, Giuseppe and Troyer, Matthias},
  title   = {Solving the quantum many-body problem with artificial neural networks},
  journal = {Science},
  volume  = {355},
  pages   = {602--606},
  year    = {2017}
}

@article{verresen2021prediction,
  author  = {Verresen, Ruben and Lukin, Mikhail D. and Vishwanath, Ashvin},
  title   = {Prediction of Toric Code Topological Order from {R}ydberg Blockade},
  journal = {Physical Review X},
  volume  = {11},
  pages   = {031005},
  year    = {2021}
}

@article{dennis2002topological,
  author  = {Dennis, Eric and Kitaev, Alexei and Landahl, Andrew and Preskill, John},
  title   = {Topological quantum memory},
  journal = {Journal of Mathematical Physics},
  volume  = {43},
  pages   = {4452--4505},
  year    = {2002}
}

@article{bravyi2014efficient,
  author  = {Bravyi, Sergey and Suchara, Martin and Vargo, Alexander},
  title   = {Efficient algorithms for maximum likelihood decoding in the surface code},
  journal = {Physical Review A},
  volume  = {90},
  pages   = {032326},
  year    = {2014}
}

@article{carrasquilla2017machine,
  author  = {Carrasquilla, Juan and Melko, Roger G.},
  title   = {Machine learning phases of matter},
  journal = {Nature Physics},
  volume  = {13},
  pages   = {431--434},
  year    = {2017},
  doi     = {10.1038/nphys4035}
}

@article{torlai2018neural,
  author  = {Torlai, Giacomo and Mazzola, Guglielmo and Carrasquilla, Juan and Troyer, Matthias and Melko, Roger and Carleo, Giuseppe},
  title   = {Neural-network quantum state tomography},
  journal = {Nature Physics},
  volume  = {14},
  pages   = {447--450},
  year    = {2018},
  doi     = {10.1038/s41567-018-0048-5}
}

@article{chen2024empowering,
  author  = {Chen, Ao and Heyl, Markus},
  title   = {Empowering deep neural quantum states through efficient optimization},
  journal = {Nature Physics},
  volume  = {20},
  pages   = {1476--1481},
  year    = {2024},
  doi     = {10.1038/s41567-024-02566-1}
}

\end{document}